\documentclass[aip,apl,numerical,preprint,superscriptaddress]{revtex4}
\usepackage{hyperref}
\usepackage{graphicx,graphics}
\usepackage{wasysym}
\usepackage{csquotes}

\begin{document}
\bibliographystyle{aipnum4-1}

\title{Reversal process of a magnetic vortex core under the combined action of a perpendicular field and spin transfer torque}

\author{N. Locatelli}
\affiliation{Unit\'e Mixte de Physique CNRS/Thales and Universit\'e Paris Sud 11, 1 av A. Fresnel, 91767 Palaiseau, France}

\author{A.E. Ekomasov}
\affiliation{Bashkir State University, Ufa, Russia}

\author{A. V. Khvalkovskiy}
\altaffiliation{Current Address: Samsung Electronics, Semiconductor R\&D (Grandis), San Jose, CA, USA}
\affiliation{A.M. Prokhorov General Physics Institute of RAS, Moscow, Russia}

\author{Sh. A. Azamatov}
\affiliation{Bashkir State University, Ufa, Russia}

\author{K.A.  Zvezdin}
\affiliation{A.M. Prokhorov General Physics Institute of RAS, Moscow, Russia}

\author{J. Grollier}
\affiliation{Unit\'e Mixte de Physique CNRS/Thales and Universit\'e Paris Sud 11, 1 av A. Fresnel, 91767 Palaiseau, France}

\author{E.G. Ekomasov}
\affiliation{Bashkir State University, Ufa, Russia}

\author{V. Cros}
\altaffiliation{Corresponding author : vincent.cros@thalesgroup.com}
\affiliation{Unit\'e Mixte de Physique CNRS/Thales and Universit\'e Paris Sud 11, 1 av A. Fresnel, 91767 Palaiseau, France}

\date{\today}

\begin{abstract}

In a nanopillar with dipolarly coupled vortices, we present an experimental and simulation study to understand how the interplay between the bias field and spin transfer torque impacts reversal of the vortex cores. We find that, depending on the current values, vortex cores might experience different physical mechanisms for their reversal, namely a static or a dynamic switching. We believe that our results might be useful in the context of vortex based non volatile memories, as a current controlled selective core switching is proposed . 

\end{abstract}

\maketitle

	Remarkable static and dynamical properties of magnetic vortices, and particularly the one of its cores ~\cite{JNN_8_Guslienko_2008,JMMM_240_Okuno_2002} has recently motivated numerous experimental and/or theoretical studies to investigate the behavior of a single vortex under the influence of current induced torques. Notably spin torques associated to rf currents have been proved to be an efficient way to reverse a vortex core. As for the vortex dynamics, very large amplitude of gyrotropic motion of vortex core can be obtained due to spin transfer torques, making vortex basic systems very interesting for the investigation of fundamental aspects of vortex dynamics~\cite{PRL_100_Bolte_2008,PRB_86_Dussaux_2012} and even chaotic behaviors~\cite{NP_8_Petit-watelot_2012} but also for their implementation as highly integrated local rf oscillators in telecommunication systems~\cite{NC_1_Dussaux_2010}.

One of the most interesting properties of magnetic vortices are their high stability in geometrically confined systems, explaining why these magnetic objects are good candidates as a new media for non-volatile magnetic memories in which information is stored either in their core polarity or chirality~\cite{NM_6_Cowburn_2007,APL_96_Choi_2010}. Thus it is of major interest to identify the physical mechanisms, often related to the static and dynamical magnetic properties of the material, that are controlling the core reversal processes. These issues are primordial for the optimization of the writing procedure but also for increasing the final speed of these vortex based memories. First studies focused on static field induced core switching~\cite{JAP_90_Kikuchi_2001,JMMM_240_Okuno_2002}, in which the vortex core breaks down and form back with opposite polarity. This mechanism leads to relatively large switching fields that are mostly material dependent (and thus not easy to modify). Recently, resonant core reversal under the presence of either an rf field~\cite{NC_2_Kammerer_2011, JAP_102_Kravchuk_2007, NP_7_Pigeau_2010, N_444_Waeyenberge_2006} or an rf current~\cite{APL_91_Kim_2007, NM_6_Yamada_2007} have been proposed to lower their switching fields but also in the perspective of memory devices, to provide a selective tool to reverse a single core. Finally, core reversal have been predicted using injection of dc current and the spin transfer torque thereof~\cite{APL_96_Choi_2010, APL_91_Sheka_2007,IJQC_110_Gaididei_2010, PRB_80_Khvalkovskiy_2009}.
	
Our main objective is to investigate how, in a system with two coupled vortices, different core reversal processes can coexist, namely static reversal at low dc currents and dynamic reversal at large currents, and more importantly how the current induced dynamics of the coupled vortices can allow to switch between the two processes. In our system i.e. a spin valve nanopillar with one vortex in each of the two magnetic layers, we have performed a detailed experimental study supported by micromagnetic simulations to understand the evolution of the core switching fields under the action of a large dc current ($I_{dc}$) and a perpendicular applied magnetic field ($H_{perp}$). The sample stack is Cu(60nm)/Py(15nm)/Cu(10nm)/Py(4nm)/Au(25nm) where Py=Ni$_{80}$Fe$_{20}$, in which circular nanopillars of two different diameters, $120$nm or $200$nm,  have been patterned by standard e-beam and ion etching lithography process. Such spin transfer oscillators based on coupled vortex dynamics have demonstrated a record linewidth in the tens of kHz range even at zero field, making them very promising for the real development of spin transfer rf devices~\cite{APL_98_Locatelli_2011, JPDAP_44_Sluka_2011}.

Our experimental procedure is to sweep the perpendicular applied field $H_{perp}$ from large negative to large positive value at different injected dc currents $I_{dc}$. In the starting configuration, the two vortices have their core polarities oriented parallel (P) and negative, and with identical chiralities (imposed by the Oersted field symmetry). This experimental procedure was applied for both $\diameter120nm$ and $\diameter200nm$ circular pillars. 
We also performed micromagnetic simulations \footnote{These simulations are done by numerical integration of the Landau-Lifshitz-Gilbert-Slonczewski equation using the micromagnetic code SpinPM based on the forth order Runge-Kutta method with an adaptive time-step control for the time integration. We use the following magnetic parameters extracted from experiments~\cite{PRB_84_Naletov_2011} : $M_{S_{thick}}=700emu/cm^{3}$ for the thick 15nm layer, and $M_{S_{thin}}=600emu/cm^3$ for the thin 4nm layer (in accordance to the FMR data for these layers), exchange stiffness $A_{thick}=1.2\mu erg/cm$ and $A_{thin}=1.1\mu erg/cm$, and Gilbert damping $\alpha=0.01$. The mesh cell size is $2\times2\times5~nm^{3}$ for the thick layer, $2\times2\times4~nm^{3}$ for the thin layer. The spin polarization of the current  is valued at $P=0.1$.} on the $\diameter200nm$ nanopillar. The simulations take into account the fact that both magnetic layers in the stack can act simultaneously as a spin polarizer and as a free layer when considering spin torque action. The current flow is assumed to be uniform through the pillar and the Oersted field is taken into account. A constant small in-plane field $H_{x}$=50~Oe is introduced to break the system's symmetry, but is not impacting the gyrotropic frequencies or any other dynamical properties. For each field and current value, we calculated the vortex dynamics of the nanopillar, with the initial state corresponding to the quiescent state at this field (i.e. with no current).

\begin{figure}[t]
  \includegraphics[width=8.5cm]{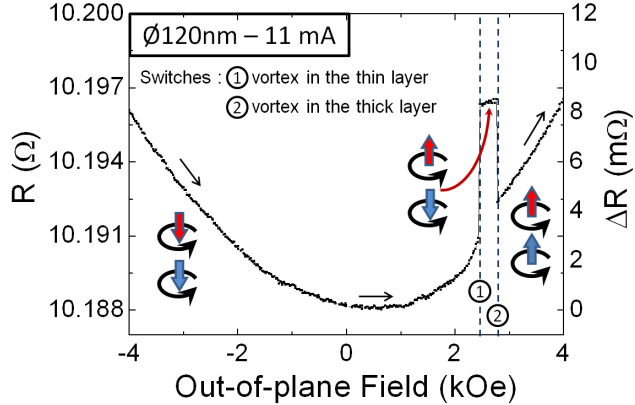}
  \caption{Resistance versus $H_{perp}$ at $I_{dc}=11mA$ measured on  $\diameter 120nm$ nanopillar. Vertical dashed lines indicate switching events, and vortices polarities are sketched for each region: top (red) correspond to thin layer's and bottom (blue) to thick layer's. }
  \label{fig.RHperp120nm11mA}
\end{figure}

	In Fig.~\ref{fig.RHperp120nm11mA}, we show a typical R($H_{perp}$) curve (measured on the 120nm diameter sample at $I_{dc}$=11mA), that describes how the core switching fields are precisely detected. At $H_{perp}$=-4kOe, the two vortex polarities are parallel and oriented downwards in our field convention ($P_{thin}=P_{thick}=-1$). Then we sweep the field towards positive values. In this field range, a  smooth parabolic variation of R is observed that is not related to a change of the vortex core configuration but rather attributed to the reversible variation with $H_{perp}$ of the out-of-plane magnetization component in the two vortex tails. The first direct evidence of a core switching event in Fig. \ref{fig.RHperp120nm11mA} is observed for $H_{perp}$=2.6kOe at which a sharp resistance variation occurs. We recently proved ~\cite{APL_98_Locatelli_2011} that this resistance change comes from the fact that a large amplitude spin transfer gyrotropic mode is generated when the two vortex polarities are AP, increasing the core-core distance and subsequently the mean resistance. In such configuration with AP polarities, a large rf peak can be detected with a frequency in the MHz range corresponding to the gyrotropic mode of the coupled vortices, typically $f$=1.1GHz at $I_{dc}$=11mA for $\diameter120nm$, and $f$=720MHz at $I_{dc}$=20mA for $\diameter200nm$. An important feature of this vortex dynamics is that the main characteristics of the excited coupled mode are close to the one of the isolated vortex in the thick layer. Consequently, the gyration of thin layer's vortex core will be neglected in a first approximation.
	
	We emphasize that two sources of spin polarization can promote large amplitude spin transfer excitations~\cite{arXiv_Sluka_2011} of the vortex core in the thick layer. A first one is due to the perpendicular spin polarization associated to the  thin layer's vortex magnetization that develops a perpendicular component under the application of the perpendicular bias field~\cite{PRL_99_Ivanov_2007, PRB_80_Khvalkovskiy_2009}. The second one is associated to the in-plane circular magnetization distribution of the thin layer's vortex that can equally contribute to the vortex excitation~\cite{APL_96_Khvalkovskiy_2010}. A precise description of the evolution of these two contributions with bias field is out of the scope of this paper. However we can notice that they can add up and thus increase the gyration amplitude of the core when the magnetic field increases in the direction of the thick layer's core.
A mean to identify which core has reversed is to measure the slope of the linear evolution of the frequency with $H_{perp}$, $f(H_{perp})$, because it is directly related to its core polarity sign : $P_{thick}=sign(df/dH_{perp})$~\cite{PRL_102_Deloubens_2009}. This feature allows us to ensure that the first resistance jump in Fig.\ref{fig.RHperp120nm11mA} is related to the switching of the vortex core in the thin layer. In the field range where the two polarities are AP, the resistance remains on this plateau with large values, until it decreases sharply back to the low resistance level at $H_{perp}$=3.1kOe, at which the core direction of the vortex in the thick Py layer eventually reverses. Note that each core switching is an irreversible process. We have repeated such R($H_{perp}$) curve for $I_{dc}$ going from 10 to 17mA. In addition, we have performed similar measurements for a $\diameter200$nm nanopillar with $I_{dc}$ going from 14 to 40mA. For each curve, the values of the two switching fields are extracted and linked either to the thin or the thick layer's vortex core switch.

\begin{figure}
  \includegraphics[width=8.5cm]{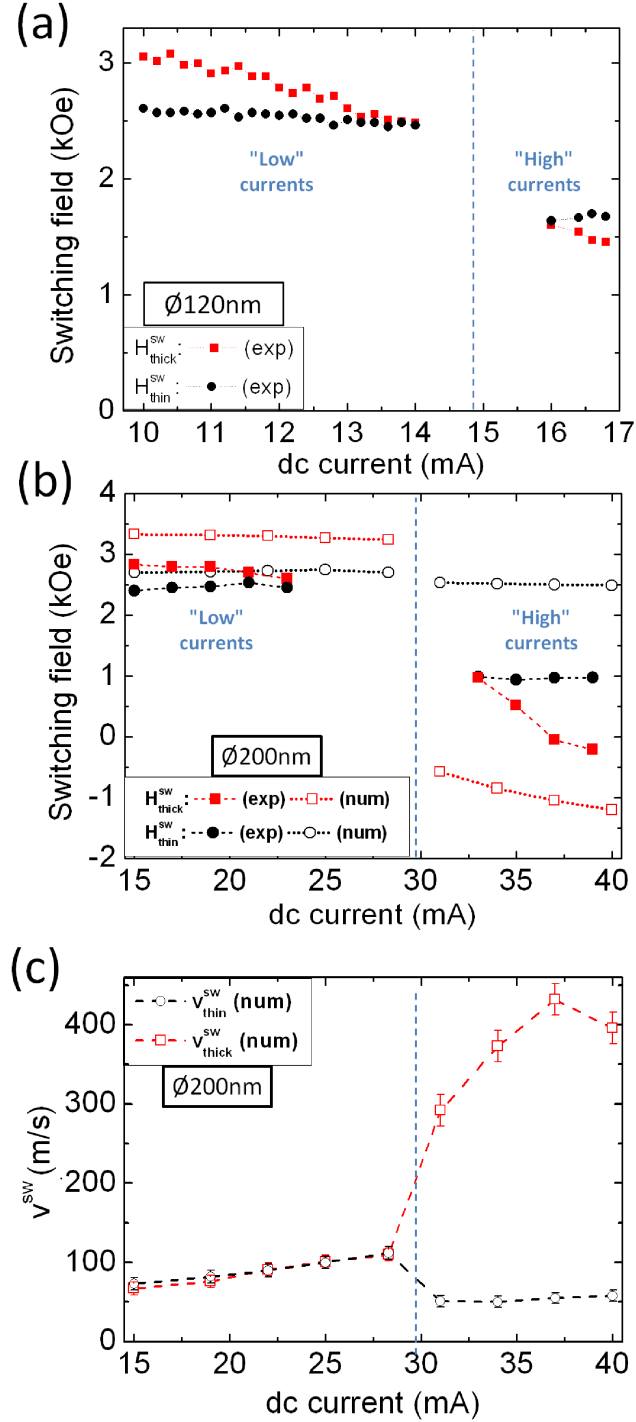}
  \caption{(a) and (b) : Experimental switching fields of the thin layer's (solid black circles) and the thick layer's (solid red squares) vortex cores (a) for the $\diameter 120nm$ pillar, and (b) for the $\diameter 200nm$ pillar, as a function of $I_{dc}$. (b) : Switching fields of the vortex cores for the $\diameter 200nm$ pillar obtained from micromagnetic simulations (open symbols). (c) Corresponding simulated velocities of the vortex cores prior to the switching events for the thin (black circles) and thick (red squares) layers'.}
  \label{fig.Hswitch}
\end{figure}
		
	In Fig \ref{fig.Hswitch}(a) and (b), we display the evolution of the two switching fields as a function of $I_{dc}$ for each pillar diameter, showing very similar trends for both nanopillars. Indeed, the variation of the two switching fields, $H_{thin}^{sw}$ and $H_{thick}^{sw}$ with $I_{dc}$ can be divided in two regimes. First, at \enquote{low} currents ($I_{dc}\leq14$mA for $\diameter120$nm and $I_{dc}\leq23$mA for $\diameter200$nm), the thin layer's vortex core is found to switch first for both diameters. Furthermore, the switching field $H_{thin}^{sw}$ is almost constant on the whole low current range, whereas $H_{thick}^{sw}$ decreases significantly to reach the values of $H_{thin}^{sw}$ at the end of this first region. As we will discuss later, these features are related to static core switching. In the second regime, at \enquote{high} currents ($I_{dc}\geq16$mA for $\diameter120$nm and $I_{dc}\geq33$mA for $\diameter200$nm), we find an opposite behavior, as the switching field of the thick layer's vortex core is always found to be smaller than the one of the thin layer's vortex. This is indeed indicative of the influence of the spin transfer induced vortex dynamics on the core reversal process. Moreover, $H_{thick}^{sw}$ decreases strongly with $I_{dc}$, passing through zero-field value and reaching negative field values for $I_{dc}\geq37$mA in the nanopillar with $\diameter200$nm (see Fig~\ref{fig.Hswitch}(b)). This is a direct indication of the dynamic switching of the vortex core, as we discuss in more details below. On the contrary, we notice that $H_{thin}^{sw}$ remains constant in the whole large current range but with a mean value strongly reduced compared to the one in \enquote{low} currents.  Between these two current regimes, we find a range where no change of resistance in R(H) curves is detected, indicating that both vortex are switching simultaneously at the timescale of the measurement. Indeed, the two switching fields are probably close enough so that any switch of one of them immediately triggers the switching of the second one.

	 In Fig.~\ref{fig.Hswitch}(b), we present the simulation results for the evolution of the two core switching fields as a function of $I_{dc}$. Depending on the field and current combinations, either static or dynamic switching of the thick layer's vortex core might exist whereas only static switching of the thin layer's core is predicted (it is explained below how to differentiate between static and dynamic switching events in the simulations). We find in these simulations that both cores experience a static switching at low $I_{dc}$, and the dependence of the switching fields for both vortices on $I_{dc}$ is very weak. At large $I_{dc}$, we find the static core switching of the vortex in the thin layer and the dynamic core switching of the vortex in the thick layer. For this latter case, the switching field $H_{thick}^{sw}$ decreases strongly with $I_{dc}$ and it even becomes negative. All these observations are in remarkable agreement with the experimental data.  In the simulations, the two vortices' cores are always found to switch independently (no \enquote{middle} current range), however this may be related to the absence of structural defects and grains in the simulated systems and to the fact that the simulations are done at T = 0K.  In Fig~\ref{fig.Hswitch}(c), we show the velocities of the cores prior to their reversal extracted from the simulations \footnote{This velocity is defined as the velocity at a field 50 Oe smaller than the static switching field}. 

Recently, several studies based on micromagnetic simulations emphasized important differences between the static and dynamic switching processes. For example, Thiaville \textit{et al.} ~\cite{PRB_67_Thiaville_2003} have shown that the \enquote{static} core reversal is mediated by a formation and propagation of a Bloch point (magnetic monopole) within the center of the vortex core. In the continuous micromagnetic limit, the switching field, which is the field at which the Bloch point is introduced into the core, is infinite. There is no precise numerical solution of this problem; the apparent field of the static core reversal yielded by the simulations is mesh-dependent, thus has a vague physical meaning. However, as was also shown in the Ref. ~\cite{PRB_67_Thiaville_2003}, the energy barrier for the core reversal reduces with an external applied field. This implies that the thermal fluctuations will result in the static core reversal when the barrier becomes comparable with the thermal energy ($k_B$T). The second process, that we referred to as \enquote{dynamical} core reversal, takes place through the deformation of the vortex shape because of the onset of a dynamical gyrofield oriented oppositely to the initial core polarity. Consequently, a vortex-antivortex pair is nucleated leading to the annihilation of the original vortex and eventually the presence of a single vortex with opposite core polarity~\cite{ N_444_Waeyenberge_2006, IJQC_110_Gaididei_2010,PRL_100_Guslienko_2008, JPCS_303_Gliga_2011}. The onset of this process (formation of the vortex-antivortex pair) can be correctly accounted for by the simulations, moreover leading to an accurate prediction of the critical velocity~\cite{NP_5_Vansteenkiste_2009} . Note that the critical velocity for the dynamical core reversal is independent on the size or shape of the sample~\cite{PRL_98_Hertel_2007}, but only on the material parameters and perpendicular bias field. At zero perpendicular bias field, the critical velocity is given by ~\cite{PRL_101_Lee_2008}: $v_{cri}\approx1.66\gamma \sqrt{A}$, valued at $306$m/s and $320$m/s for the thin and thick layer's respectively. When the magnetic bias field $H_{perp}$ is applied, the critical velocity changes approximately linearly with the field~\cite{APL_96_Khvalkovskiy_2010_2}. 

These facts provide us with a deeper insight into relationship between simulations and experimental data. In the \enquote{low} currents range, the spin transfer induced dynamics are associated with small orbit radius, and hence small velocity. As a consequence, the condition that the vortex core of the thick layer reaches its critical velocity is never fulfilled for any current-field combination. The two reversals then correspond to core breakdown, starting in the core center without formation of a vortex-antivortex pair. Indeed, a potential dynamic reversal at the same field would require much larger velocities that the actual ones reached by the cores prior to their static switching (see Fig~\ref{fig.Hswitch}(c)). \footnote{Additional simulations with a finer mesh (in-plane mesh step of 1.8 nm) and very large pulses of polarization independent on both layers showed that the vortex cores are stable at least 150 m/s at the same fields.}. Moreover, we find that, at any field value, the core profile is narrower for the thin layer's vortex. Consequently, it results that the energy barrier for the static reversal is smaller for this layer, making it easier to switch by thermal fluctuations ~\cite{ PRB_67_Thiaville_2003}. In simulations, the static switching fields are independent of $I_{dc}$. We believe that the experimental decrease of $H_{thick}^{sw}$ with $I_{dc}$ (see Fig. \ref{fig.Hswitch}(a) and (b)) is dueto its finite velocity before the switching, that leads to a narrowing of the vortex core. This feature reduces the energy barrier for the switching while the thermal fluctuations increase because of the heating effect by the current. 

For \enquote{high} currents, while sweeping the field the system reaches a point at which the evolution of the thick layer's core velocity crosses the critical velocity for a given $H_{perp}$ before the vortex core breakdown field $H_{thin}^{static}$ is reached. Therefore, its core eventually switches before the thin layer's one. As shown in Fig~\ref{fig.Hswitch}(c), the reversals of the thick layer's core then happens at velocities that strongly increase when $I_{dc}$ increases, corresponding to the decrease of the observed switching field with $I_{dc}$, in agreement with predictions by A.V. Khvalkovskiy et al~\cite{APL_96_Khvalkovskiy_2010_2}. Interestingly, this core dynamical switching can happen even for negative field values, i.e. an applied field oriented in the direction of the initial core polarity. This is then an experimental confirmation that spin transfer core switching induced solely by a dc current is possible even at zero field. 
In the high-current regime, when the field is increased further, the \enquote{static} critical switching field of the thin layer's core is eventually reached. Similarly to the \enquote{low} current regime, the switching field of the thin layer's vortex due this \enquote{static} process is almost independent of $I_{dc}$. Notably, we find different values for the static switching field of the thin layer's vortex between the two current regimes in the experiments. This large difference might be related to the dipolar interaction between the two cores: in the \enquote{low} current region, the thin layer's core switches being in the (P) configuration, whereas in the \enquote{high} current region, the thin layer's core switches being in the (AP) configuration. 
	 In conclusion, we present a combined experimental and simulation study of the variation of the switching fields of vortex cores as a function of an injected dc current in a spin valve system containing two vortices, one in each magnetic layer having different thicknesses. We identify the mechanisms responsible for the core reversal under the combined action of spin transfer torque and perpendicular field opposite to the initial core polarity. At low currents, both cores reversals occur through a static reversal of the vortex core. In this regime, the switching fields are defined by the intrinsic material parameters and their dependence with $I_{dc}$ is weak.. At large current, the large amplitude gyrotropic motion induced by spin transfer allows for the thick layer's vortex to reach the critical velocity for a dynamical reversal, mediated by the creation of a pair of vortex-antivortex. In this regime the switching field of the vortex core has a significant dependence on the current. Our results are thus interesting in the perspective of development of non volatile vortex based memories as we demonstrate the possibility to obtain a selective switching event in a two-vortices based devices, simply through the application of a combination of dc current and pulsed field.

	Financial support by the ANR agency (VOICE PNANO-09-P231-36 and SPINNOVA 11-NANO-0016), EU grant MASTER No. NMP-FP7-212257, RFBR grant 10-02-01162 is acknowledged.

\bibliography{Biblio}

\end{document}